\documentclass[onecolumn,aps,preprint,showpacs,amsmath,amssymb]{revtex4}
\usepackage{graphicx}
\usepackage{latexsym}
\usepackage{dcolumn}
\usepackage{stmaryrd}

\usepackage{bm}

\begin{document}
\newcommand{\eq}{\begin{equation}}                                                                         
\newcommand{\eqe}{\end{equation}}             

\title{Generalized characteristics of the homogenous magneto hydrodynamical equations} 

\author{ I. F. Barna}
\address{Atomic Energy Research Institute of the Hungarian Academy 
of Sciences, \\ (KFKI-AEKI), H-1525 Budapest, P.O. Box 49, Hungary \\ 
Email: barnai@sunserv.kfki.hu}
\date{\today}

%\maketitle
%%%%%%%%%%%%%%%%%%%%%%%%%%%%%%%%%%%%%%%%%%%%%%%%%%%%%%%%%%%%%%%%%%%%%%%
\begin{abstract}
With the help of the generalized characteristics(GC) of the first order partial 
differential equations(PDE) we calculate the differential equation system 
of characteristics of the homogenous magneto hydrodynamical equations(MHD). 
%We analyze the restricted one dimensional motion, calculate the fix points, 
%check the stability of the system and try to use the Painleve test. 

\end{abstract}
%\ead{barnai@sunserv.kfki.hu}%\affiliation{ 

%\draft
\maketitle
%%%%%%%%%%%%%%%%%%%%%%%%%%%%%%%%%%%%%%%%%%%%%%%%%%%%%%%%%%%%%%%%%%%%%%%                                    
%\widetext 
%\begin{multicols}{2}
%%%%%%%%%%%%%%%%%%%%%%%%%%%%%%%%%%%%%%%%%%%%%%%%%%%%%%%%%%%%%%%%%%%%%%%                                    

\section{Introduction}
The investigation of the magneto hydrodynamical(MHD) equations is crucial 
importance to understand plasma instabilities in the future fusion facilities like ITER. 
Nowadays large efforts are made to numerically solve the complete MHD equations 
on a torus or on a more realistic stellarator geometry. \cite{powell} 
However, simplified models of the complete MHD equation may help us to get a deeper 
understanding physical mechanism of plasma or let more insight into the structure of magnetized fluids. 
In the following study we shortly introduce the mathematical formalism of the generalized 
characteristics of first order partial differential equations and apply this theory to the 
full three dimensional MHD equations.

%%%%%%%%%%%%%%%%%%%%%%%%%%%%%%%%%%%%%%%%%%%%%%%%%%%%%%%%%%%%%%%%%%%%%%%%%%%%%                              
\section{Theory}       
It is well known from the theory of first order linear PDEs that along the characteristic line 
the original PDE becomes an ordinary differential equation(ODE). Therefore, we can make an 
important observation, along the characteristics the solution 
is constant. For linear equations the characteristic curve is a line. For one space and time 
dimension the characteristic lines give us a qualitative picture about the solution of the 
equation on the plain. The proof of this statement can be found in any textbook \cite{difk}

It the following we briefly introduce the mathematics of the 
generalized characteristics of the fist order PDEs. 
Let's consider the following first order PDE, which can be non-linear as well:  

\eq
F(x,t,u(x,t), p ,q) = 0
\eqe
we use the standard notation of  $ p = \partial u(x,t)/\partial x$, $ q = \partial u(x,t)/\partial t$. 
According to the book of Melikyan \cite{arik} the differential equation system of the characteristics 
is the following: 
\begin{eqnarray}                                                                                          
 \dot{x} = F_p, \hspace*{1cm}
 \dot{t} = F_q,   \\
 \dot{u}(x,t) = p \cdot F_p + q \cdot F_q,  \\
 \dot{p}=  - F_x + p \cdot F_u, \hspace*{1cm}
 \dot{q}=  - F_t + q \cdot F_u  
  \end{eqnarray}                                                                                           
where  $\centerdot = \frac{d}{d \tau} $.  
To avoid further confusions and misunderstanding we  
use the more detailed notation: 
\begin{center}                                                                                          
\begin{eqnarray}
 \frac{dx}{d\tau} =  \frac{\partial F}{\partial 
 (\frac{\partial u}{\partial x}) }, \hspace*{1cm}
 \frac{dt}{d\tau} =  \frac{\partial F}{\partial 
 (\frac{\partial u}{\partial t}) }, \\
 \frac{du}{d\tau} = \frac{\partial u}{\partial x} \cdot \frac{\partial F}{\partial 
 (\frac{\partial u}{\partial x}) } +
 \frac{\partial u}{\partial t} \cdot \frac{\partial F}{\partial (\frac{\partial u}{\partial t}) }, \\
 \frac{dp}{d\tau} = -\frac{\partial F}{\partial x}  -
 \frac{\partial u}{\partial x} \cdot \frac{\partial F}{\partial u },  \hspace*{1cm}
 \frac{dq}{d\tau} = -\frac{\partial F}{\partial t}  -
 \frac{\partial u}{\partial t} \cdot \frac{\partial F}{\partial u } \\
\end{eqnarray}    
\end{center}                                                                                          

  These equations can be easily generalized to a higher dimensional equation system
  $i=1...n$ as well.
  
\eq
F_i(x,y,z,t,u_i(x,y,z,t),p_i,q_i,r_i,s_i) = 0
\eqe
with the notation of  $ p_i = \partial u_i(x,y,z,t)/\partial x $ , 
$ q_i = \partial u_i(x,y,z,t)/\partial y$, 
$ r_i = \partial u_i(x,y,z,t)/\partial z $, 
$ s_i = \partial u_I(x,y,z,t)/\partial t$. 
with the above given detailed notation 

\begin{eqnarray}                                                                                          
 \frac{dx}{d\tau} =  \frac{\partial F_i}{\partial 
 (\frac{\partial u_i}{\partial x}) } \nonumber \\
 \frac{dy}{d\tau} =  \frac{\partial F_i}{\partial 
 (\frac{\partial u_i}{\partial y}) } \nonumber \\
 \frac{dz}{d\tau} =  \frac{\partial F_i}{\partial 
 (\frac{\partial u_i}{\partial z}) } \nonumber \\
 \frac{dt}{d\tau} =  \frac{\partial F_i}{\partial 
 (\frac{\partial u_i}{\partial t}) } \nonumber \\
   \frac{du_i}{d\tau} = \frac{\partial u_i}{\partial x} \cdot \frac{\partial F_i}{\partial  
 (\frac{\partial u_i}{\partial x}) }   +
 \frac{\partial u_i}{\partial y} \cdot \frac{\partial F_i}{\partial   
 (\frac{\partial u_i}{\partial y}) } + 
 \frac{\partial u_i}{\partial z} \cdot \frac{\partial F_i}{\partial   
 (\frac{\partial u_i}{\partial z}) }  + 
 \frac{\partial u_i}{\partial t} \cdot \frac{\partial F_i}{\partial
(\frac{\partial u_i}{\partial t}) } \nonumber \\
 \frac{dp_i}{d\tau} = -\frac{\partial F_i}{\partial x}  -
 \frac{\partial u_i}{\partial x} \cdot \frac{\partial F_i}{\partial u_i } \nonumber \\
 \frac{dq_i}{d\tau} = -\frac{\partial F_i}{\partial y}  -
 \frac{\partial u_i}{\partial y} \cdot \frac{\partial F_i}{\partial u_i } \nonumber \\
 \frac{dr_i}{d\tau} = -\frac{\partial F_i}{\partial z}  -
 \frac{\partial u_i}{\partial z} \cdot \frac{\partial F_i}{\partial u_i } \nonumber \\
 \frac{ds_i}{d\tau} = -\frac{\partial F_i}{\partial t}  -
 \frac{\partial u_i}{\partial t} \cdot \frac{\partial F_i}{\partial u_i } 
 \label{gen}
 \end{eqnarray}    
These kind of equations can be applied to any kind of hyperbolic equation systems, 
like gas dynamical problems \cite{difk}, multi-phase flows \cite{wendr} or to MHD.  

The governing equations for an ideal, non-relativistic compressible 
plasma may be written in different forms if the following assumptions hold:
\eq                                                                               
  \frac{\lambda}{L} \ll 1, \hspace*{1cm}   \frac{\epsilon}{\tau \sigma} \ll 1,  \hspace*{1cm}
   \left( \frac{v}{c} \right) ^2 \ll 1, \hspace*{1cm}   \frac{\mu}{\rho V L}  \ll 1
\eqe
where $ \rho, v, \tau $ and L are, respectively, characteristic density, speed, time and length 
scales for the problem, c is the speed of light, and $\epsilon$ and $\sigma$ is the dielectric 
constant and conductivity of the fluid. 
In conservative variables, 
the governing equations, which is a combination of Euler equations 
of gas dynamics and the Maxwell equations of electromagnetics, is the following: 
\eq
\frac{\partial}{ \partial t} 
\left(  
\begin{array}{c}
\rho \\
\rho {\bf{v}}\\
{\bf{B}}\\                                                                                       
E 
 \end{array} \right) 
     + \nabla \left(  
\begin{array}{c}
\rho {\bf{v}}\\
\rho {\bf{v}}{\bf{v}} + {\bf{I}}\left(\pi + \frac{\bf{B}\bf{B}}{2} \right) -\bf{B}\bf{B}    \\
{\bf{v}}{\bf{B}}   - {\bf{B}}{\bf{v}}   \\     
\left( E + \pi +   \frac{\bf{B}\bf{B}}{2} \right){\bf{v}} - {\bf{B}} ({\bf{v}}\cdot {\bf{B}} ) 
 \end{array} \right) 
  =0 \label{mhd}
\eqe
where $\bf{I}$ is the 3 $\times $3 identity matrix, $\rho$ is the density, $\bf{v}$ is the velocity, 
$\pi$ is the pressure (to avoid confusion with the notation used for partial differential equations 
$ p = \partial u(x,t)/\partial x$), {\bf{B}} is the magnetic field, and E is the energy, defined as:
\eq 
E = \frac{\pi}{\gamma -1} + \rho\frac{ {\bf{v}}\cdot {\bf{v}}}{2} + \frac{ {\bf{B}}\cdot {\bf{B}}}{2} 
\eqe
Solution of these equations can help to understand a number of problems governed by fluid-dynamics and 
electromagnetic effects. 

Given the following primitive variables $ {\bf{w}} = (\rho, v_x, v_y, v_z, B_x, B_y, B_z, \pi )$,     
the MHD equations (\ref{mhd}) may be written in quasi-linear form as: 
\eq
 \frac{\partial \bf{w} }{\partial t}  + \underline{\underline A} \frac{\partial {\bf{w}} }{\partial x}   +  
 \underline{\underline B} \frac{\partial {\bf{w}} }{\partial y}   +  
 \underline{\underline C} \frac{\partial {\bf{w}} }{\partial z}    = 0 \label{mtr}
\eqe
where  $ \underline{\underline A}, \underline{\underline B}, \underline{\underline C} $ 
are $8 \times 8 $ matrices. 

With a 
%\begin{table}
%\caption{ $ \underline{\underline A} = $ }
\eq 
\underline{\underline A} =
 \vspace*{2mm}
  \left[ \begin{tabular}{cccccccc} 
 $  v_x $  &  $ \rho$   &  0    &  0   & 0 & 0   & 0 & 0    \\
  0   & $v_x$  &  0 &   0 & $ -\frac{B_x}{\rho} $&  $ \frac{B_y}{\rho} $ &  $ \frac{B_z}{\rho} $    & $ \frac{1}{\rho} $    \\
  0 &   0   & $v_x$   & 0  &   $ -\frac{B_y}{\rho} $ & $ -\frac{B_x}{\rho} $  & 0 &  0    \\ 
 0     &  0  & 0  & $v_x$  &  $ -\frac{B_z}{\rho} $  & 0  &  $ -\frac{B_x}{\rho} $ &  0        \\
 0  & 0   & 0  & 0  & 0 & 0 & 0   & 0  \\
 0 &  $B_y$   & $-B_x$ &  0 & $ -v_y$   & $v_x$  & 0 & 0    \\ 
 0    &  $B_z$    &  0   & $ -B_x$  & $-v_z$  & 0   & $v_x$  &  0    \\  
 0  & $ \gamma \pi $  &  0 &  0   &  $(\gamma-1)\bf{v}\cdot \bf{ B}   $ &  0  &  0 & $v_x$     \\   
\end{tabular} \right]  
\eqe

where $\gamma  $ is the compressibility of the fluid. 
Matrix $\underline{\underline A} $ is singular - the fifth is zero, leading to 
a zero eigenvalue after the diagonalization.    
\begin{eqnarray}                                                                                          
 \lambda^A_1 =  0,  \>\>\>\>
 \lambda^A_2 =  v_x,  \>\>\>\>
 \lambda^A_3 =  v_x + \frac{B_x}{\sqrt{\rho}}, \>\>\> 
 \lambda^A_4 =  v_x -  \frac{B_x}{\sqrt{\rho}}  \nonumber \\
 \lambda^A_5 =  v_x +  \frac{1}{\sqrt{2\rho}}\sqrt{{\bf{B}}^2+ \gamma \pi + \sqrt{({\bf{B}}^2+ \gamma \pi  
 )^2-4 \gamma \pi B_x^2}} \\
 \lambda^A_6 =  v_x -  \frac{1}{\sqrt{2\rho}}\sqrt{{\bf{B}}^2+ \gamma \pi+ \sqrt{({\bf{B}}^2+ \gamma \pi
 )^2-4 \gamma \pi B_x^2}} \\
 \lambda^A_7 =  v_x +  \frac{1}{\sqrt{2\rho}}\sqrt{{\bf{B}}^2+ \gamma \pi - \sqrt{({\bf{B}}^2+ \gamma \pi 
 )^2-4 \gamma \pi B_x^2}} \\ 
 \lambda^A_8 =  v_x -  \frac{1}{\sqrt{2\rho}}\sqrt{{\bf{B}}^2+ \gamma \pi - \sqrt{({\bf{B}}^2+ \gamma \pi 
 )^2-4 \gamma \pi B_x^2}}  
\end{eqnarray}    

The eigenvalues of the system are well known, and they correspond to: \\
$\bullet$ one entropy wave $ \lambda^A_2 $ traveling with speed $v_x$,\\
$\bullet$ two Alfv\'en waves $ \lambda^A_{2,3} $ traveling with speed  $ v_x \pm c_c $ where 
$ c_a = \frac{B_x}{\sqrt{\rho}}$  is the Alfv\'en speed,\\
$\bullet$ four magneto-acoustic waves ( $\lambda^A_5 ...  \lambda^A_8$ )

For completeness we calculate and present the eigenvalues for the y and z direction too. 
Matrices $ \underline{\underline B}, \underline{\underline C} $ and the corresponding eigenvalues 
are very similar to $ \underline{\underline A} $ and have the same structure. 

\eq 
\underline{\underline B} =
 \vspace*{2mm} \left[
\begin{tabular}{cccccccc} 
   $v_y$   &  0  &  $ \rho$     &  0   & 0 & 0   & 0 & 0    \\
  0   & $v_y$  &  0 &   0 & $ -\frac{B_y}{\rho} $&  $ -\frac{B_x}{\rho} $ & 0   & 0    \\
  0 &   0   & $v_y$   & 0  &   $ \frac{B_x}{\rho} $ & $ -\frac{B_y}{\rho} $  & $\frac{B_z}{\rho} $  &  $\frac{1}{\rho}$    \\ 
 0     &  0  & 0  & $v_y$  &  0  & $-\frac{B_z}{\rho} $  &  $ -\frac{B_y}{\rho} $ &  0        \\
 0  & $-B_y$   & $-B_x$  & 0  & $v_y$ & $-v_x$ & 0   & 0  \\
 0 &  0   & 0 &  0 &  0   & 0  & 0 & 0    \\ 
 0    &  0    &  $B_z$   & $ -B_y$  & 0  & $v_z$   & $v_y$  &  0    \\  
 0  &  0    & $ \gamma \pi$ &  0   & 0 & $(\gamma-1)\bf{v}\cdot \bf{ B}   $   &  0 & $v_y$     \\   
\end{tabular}   \right]                                             
\eqe

\begin{eqnarray}                                                                                          
 \lambda^B_1 =  0,  \>\>\>\>
 \lambda^B_2 =  v_y,  \>\>\>\>
 \lambda^B_3 =  v_y + \frac{B_y}{\sqrt{\rho}}, \>\>\> 
 \lambda^B_4 =  v_y -  \frac{B_y}{\sqrt{\rho}}  \nonumber \\
 \lambda^B_5 =  v_y +  \frac{1}{\sqrt{2\rho}}\sqrt{{\bf{B}}^2+\gamma \pi+ \sqrt{({\bf{B}}^2+ 
 \gamma \pi )^2-4 \gamma \pi B_y^2}} \\
 \lambda^B_6 =  v_y -  \frac{1}{\sqrt{2\rho}}\sqrt{{\bf{B}}^2+\gamma \pi+ \sqrt{({\bf{B}}^2+ 
 \gamma \pi )^2-4 \gamma \pi B_y^2}} \\
 \lambda^B_7 =  v_y +  \frac{1}{\sqrt{2\rho}}\sqrt{{\bf{B}}^2+\gamma \pi - \sqrt{({\bf{B}}^2+ 
 \gamma \pi )^2-4 \gamma \pi B_y^2}} \\ 
 \lambda^B_8 =  v_y -  \frac{1}{\sqrt{2\rho}}\sqrt{{\bf{B}}^2+ \gamma \pi - \sqrt{({\bf{B}}^2+ 
 \gamma \pi )^2-4 \gamma \pi B_y^2}}  
\end{eqnarray}    

\eq 
\underline{\underline C} =
 \vspace*{2mm}  \left[
\begin{tabular}{cccccccc} 
   $v_z$   &  0  &    0   &  $ \rho$   & 0 & 0   & 0 & 0    \\
  0   &  $v_z$  &  0 &   0 & $ -\frac{B_z}{\rho} $& 0 &  $ -\frac{B_x}{\rho} $   & 0    \\
  0 &   0   &  $v_z$   & 0  & 0  & $ -\frac{B_z}{\rho} $  &    $ -\frac{B_y}{\rho} $ &  0    \\ 
 0     &  0  & 0  &  $v_z$  &  $ \frac{B_x}{\rho} $  &   $ \frac{B_y}{\rho} $ & 
  $ -\frac{B_z}{\rho} $ &   $ \frac{1}{\rho} $   \\
 0 &  $-B_z$   & 0 &  $B_x$ &  $v_z$    & 0  &  $-v_x$ & 0    \\ 
 0  & 0 &  $-B_z$   & $ B_y$  & 0  &  $v_z$  &  $-v_y$  &  0    \\  
 0  & 0 & 0  & 0  & 0 & 0 & 0   & 0  \\
 0  &  0    & 0 &  $\gamma \pi$   & 0  &  0  & $(g-1)\bf{v} \bf{ B}  $  &  $v_z$     \\   
\end{tabular}     \right]                                            
\eqe

\begin{eqnarray}                                                                                          
 \lambda^C_1 =  0,  \>\>\>\>
 \lambda^C_2 =  v_z,  \>\>\>\>
 \lambda^C_3 =  v_z + \frac{B_z}{\sqrt{\rho}}, \>\>\> 
 \lambda^C_4 =  v_z -  \frac{B_z}{\sqrt{\rho}}  \nonumber \\
 \lambda^C_5 =  v_z+  \frac{1}{\sqrt{2\rho}}\sqrt{{\bf{B}}^2+ \gamma \pi+ \sqrt{({\bf{B}}^2+ \gamma \pi 
 )^2-4 \gamma \pi B_z^2}} \\
 \lambda^C_6 =  v_z -  \frac{1}{\sqrt{2\rho}}\sqrt{{\bf{B}}^2+ \gamma \pi+ \sqrt{({\bf{B}}^2+ \gamma \pi
  )^2-4 \gamma \pi B_z^2}} \\
 \lambda^C_7 =  v_z +  \frac{1}{\sqrt{2\rho}}\sqrt{{\bf{B}}^2+\gamma \pi - \sqrt{({\bf{B}}^2+ \gamma \pi
  )^2-4 \gamma \pi B_z^2}} \\ 
 \lambda^C_8 =  v_z -  \frac{1}{\sqrt{2\rho}}\sqrt{{\bf{B}}^2+\gamma \pi - \sqrt{({\bf{B}}^2+ \gamma \pi
  )^2-4 \gamma \pi B_z^2}}  
\end{eqnarray}

The matrix form of the MHD equations (\ref{mtr}) reads as follows: 
\begin{eqnarray}    
\frac{\partial \rho}{\partial t} = 0  \nonumber \\
\frac{\partial v_x}{\partial t} + v_x \frac{\partial v_x}{\partial x} +
 v_y \frac{\partial v_x}{\partial y} + v_z \frac{\partial v_x}{\partial z}  = 0 \nonumber \\
\frac{\partial v_y}{\partial t} + 
\left( v_x + \frac{B_x}{\sqrt{\rho}} \right) \frac{\partial v_y}{\partial x} +
\left( v_y + \frac{B_y}{\sqrt{\rho}} \right) \frac{\partial v_y}{\partial y} + 
\left( v_z + \frac{B_z}{\sqrt{\rho}} \right) \frac{\partial v_y}{\partial z} = 0 \nonumber \\
\frac{\partial v_z}{\partial t} + 
\left( v_x - \frac{B_x}{\sqrt{\rho}} \right) \frac{\partial v_z}{\partial x} +
\left( v_y - \frac{B_y}{\sqrt{\rho}} \right) \frac{\partial v_z}{\partial y} + 
\left( v_z - \frac{B_z}{\sqrt{\rho}} \right) \frac{\partial v_z}{\partial z} = 0 \nonumber \\
\frac{\partial B_x}{\partial t} + 
\left( v_x   +  \frac{1}{\sqrt{2\rho}}\sqrt{{\bf{B}}^2+ \gamma \pi + \sqrt{({\bf{B}}^2+ \gamma \pi  
 )^2-4 \gamma \pi B_x^2}} \right) \frac{\partial B_x}{\partial x} +  \nonumber \\
\left( v_y  +  \frac{1}{\sqrt{2\rho}}\sqrt{{\bf{B}}^2+ \gamma \pi + \sqrt{({\bf{B}}^2+ \gamma \pi  
 )^2-4 \gamma \pi B_y^2}} \right) \frac{\partial B_x}{\partial y} + \nonumber \\
\left( v_z +  \frac{1}{\sqrt{2\rho}}\sqrt{{\bf{B}}^2+ \gamma \pi + \sqrt{({\bf{B}}^2+ \gamma \pi  
 )^2-4 \gamma \pi B_z^2}} \right) \frac{\partial B_x}{\partial z} = 0 \nonumber \\
\frac{\partial B_y}{\partial t} + 
\left( v_x   - \frac{1}{\sqrt{2\rho}}\sqrt{{\bf{B}}^2+ \gamma \pi + \sqrt{({\bf{B}}^2+ \gamma \pi  
 )^2-4 \gamma \pi B_x^2}} \right) \frac{\partial B_y}{\partial x} +  \nonumber \\
\left( v_y  -  \frac{1}{\sqrt{2\rho}}\sqrt{{\bf{B}}^2+ \gamma \pi + \sqrt{({\bf{B}}^2+ \gamma \pi  
 )^2-4 \gamma \pi B_y^2}} \right) \frac{\partial B_y}{\partial y} + \nonumber \\
\left( v_z -  \frac{1}{\sqrt{2\rho}}\sqrt{{\bf{B}}^2+ \gamma \pi + \sqrt{({\bf{B}}^2+ \gamma \pi  
 )^2-4 \gamma \pi B_z^2}} \right) \frac{\partial B_y}{\partial z} = 0 \nonumber \\
\frac{\partial B_z}{\partial t} + 
\left( v_x   + \frac{1}{\sqrt{2\rho}}\sqrt{{\bf{B}}^2+ \gamma \pi - \sqrt{({\bf{B}}^2+ \gamma \pi  
 )^2-4 \gamma \pi B_x^2}} \right) \frac{\partial B_z}{\partial x} +  \nonumber \\
\left( v_y  +  \frac{1}{\sqrt{2\rho}}\sqrt{{\bf{B}}^2+ \gamma \pi - \sqrt{({\bf{B}}^2+ \gamma \pi  
 )^2-4 \gamma \pi B_y^2}} \right) \frac{\partial B_z}{\partial y} + \nonumber \\
\left( v_z +  \frac{1}{\sqrt{2\rho}}\sqrt{{\bf{B}}^2+ \gamma \pi - \sqrt{({\bf{B}}^2+ \gamma \pi  
 )^2-4 \gamma \pi B_z^2}} \right) \frac{\partial B_z}{\partial z} = 0 \nonumber \\
\frac{\partial \pi}{\partial t} + 
\left( v_x   - \frac{1}{\sqrt{2\rho}}\sqrt{{\bf{B}}^2+ \gamma \pi - \sqrt{({\bf{B}}^2+ \gamma \pi  
 )^2-4 \gamma \pi B_x^2}} \right) \frac{\partial \pi}{\partial x} +  \nonumber \\
\left( v_y -  \frac{1}{\sqrt{2\rho}}\sqrt{{\bf{B}}^2+ \gamma \pi - \sqrt{({\bf{B}}^2+ \gamma \pi  
 )^2-4 \gamma \pi B_y^2}} \right) \frac{\partial \pi}{\partial y} + \nonumber \\
\left( v_z - \frac{1}{\sqrt{2\rho}}\sqrt{{\bf{B}}^2+ \gamma \pi - \sqrt{({\bf{B}}^2+ \gamma \pi  
 )^2-4 \gamma \pi B_z^2}} \right) \frac{\partial \pi}{\partial z} = 0  
\label{last}
\end{eqnarray}    
Now applying the equations of (\ref{gen}), after a tedious derivation we may get the complete equation 
system of the generalized characteristics.  \\ 
For the first equation of (\ref{last})  $ \frac{\partial \rho}{ \partial t} = 0 $ the equation system of 
characteristics is trivial $ \frac{d t}{ d \tau} = 1, \>\> \frac{d \rho}{ d \tau} = s $ where 
$\tau$ is the parameter of the characteristic curve. It is clear that time can be used as a natural 
parameter too, so the equation of the first characteristics is  $ \frac{d \rho}{ d t} = s_1 $. 
For the second equation of (\ref{last}) the system becomes much more complicated:\\  
$ \frac{dx}{ dt} = v_x, \>\> \frac{dy}{ dt} = v_y, \>\> \frac{dz}{ dt} = v_z, \>\>, 
\frac{dv_x}{ dt} = p_2v_x + q_2v_y + r_2v_z + s_2, \>\> \frac{dp_2}{ dt} = -p_2^2, \>\> \>\>, 
\frac{dq_2}{ dt} = -p_2q_2, \\ \frac{dr_2}{ dt} = -p_2r_2, \>\>  \frac{ds_2}{ dt} = -p_2s_2, \>\> \>\> $  \\ 
The equation systems of the last six variable are the following: \\ 
$ \frac{dx}{ dt} = v_x + \frac{B_x}{\sqrt{\rho}}, \>\> \frac{dy}{ dt} = v_y  + \frac{B_y}{\sqrt{\rho}} 
\>\> \frac{dz}{ dt} = v_z  + \frac{B_z}{\sqrt{\rho}}, \>\>  
\frac{d v_y}{ dt} = p_3(v_x + \frac{B_x}{\sqrt{\rho}}) + q_3(v_y + \frac{B_y}{\sqrt{\rho}}) + 
r_3(v_z + \frac{B_z}{\sqrt{\rho}}) + s_3,  \\
 \frac{d p_3}{ dt} = -p_3q_3, \>\> \frac{d q_3}{ dt} = -q_3^2, \>\> \frac{d r_3}{ dt} = -r_3q_3, \>\>   
 \frac{d s_3}{ dt} = -p_3s_3, $ \\ 
 
$ \frac{dx}{ dt} = v_x - \frac{B_x}{\sqrt{\rho}}, \>\> \frac{dy}{ dt} = v_y - \frac{B_y}{\sqrt{\rho}} 
\>\> \frac{dz}{ dt} = v_z  - \frac{B_z}{\sqrt{\rho}}, \>\>  
\frac{d v_z}{ dt} = p_4(v_x - \frac{B_x}{\sqrt{\rho}}) + q_4(v_y - \frac{B_y}{\sqrt{\rho}}) + 
r_4(v_z - \frac{B_z}{\sqrt{\rho}}) + s_4,  \\
 \frac{d p_4}{ dt} = -p_4r_4, \>\> \frac{d q_4}{ dt} = -q_4r_4, \>\> \frac{d r_4}{ dt} = -r_4^2, \>\>   
 \frac{d s_4}{ dt} = -r_4s_4, $  \\ 
 
$ \frac{dx}{ dt} = \lambda_5^A, \>\> \frac{dy}{ dt} = \lambda_5^B,  
\>\> \frac{dz}{ dt} =\lambda_5^C  , \>\>  
\frac{d B_x}{ dt} = p_5\lambda_5^A + q_5\lambda_5^B + 
r_5\lambda_5^C + s_5,  \\
 \frac{d p_5}{ dt} = -p_5\left [ \left( \frac{\partial}{\partial B_x}  \lambda_5^A \right)p_5 +
 \left( \frac{\partial}{\partial B_x}  \lambda_5^B \right)q_5 + 
 \left( \frac{\partial}{\partial B_x}  \lambda_5^C \right)r_5   
 \right], \\
  \frac{d q_5}{ dt} = -q_5\left [ \left( \frac{\partial}{\partial B_x}  \lambda_5^A \right)p_5 +
 \left( \frac{\partial}{\partial B_x}  \lambda_5^B \right)q_5 + 
 \left( \frac{\partial}{\partial B_x}  \lambda_5^C \right)r_5   
 \right],  \\
  \frac{d r_5}{ dt} =  -r_5\left [ \left( \frac{\partial}{\partial B_x}  \lambda_5^A \right)p_5 +
 \left( \frac{\partial}{\partial B_x}  \lambda_5^B \right)q_5 + 
 \left( \frac{\partial}{\partial B_x}  \lambda_5^C \right)r_5   
 \right],  \\  
 \frac{d s_5}{ dt} =  -s_5\left [ \left( \frac{\partial}{\partial B_x}  \lambda_5^A \right)p_5 +
 \left( \frac{\partial}{\partial B_x}  \lambda_5^B \right)q_5 + 
 \left( \frac{\partial}{\partial B_x}  \lambda_5^C \right)r_5   
 \right],  \\     $ \\

$ \frac{dx}{ dt} = \lambda_6^A, \>\> \frac{dy}{ dt} = \lambda_6^B,  
\>\> \frac{dz}{ dt} =\lambda_6^C  , \>\>  
\frac{d B_y}{ dt} = p_6\lambda_6^A + q_6\lambda_6^B + 
r_6\lambda_6^C + s_6,  \\
 \frac{d p_6}{ dt} = -p_6\left [ \left( \frac{\partial}{\partial B_y}  \lambda_6^A \right)p_6 +
 \left( \frac{\partial}{\partial B_y}  \lambda_6^B \right)q_6 + 
 \left( \frac{\partial}{\partial B_y}  \lambda_6^C \right)r_6   
 \right], \\
  \frac{d q_6}{ dt} = -q_6\left [ \left( \frac{\partial}{\partial B_y}  \lambda_6^A \right)p_6 +
 \left( \frac{\partial}{\partial B_y}  \lambda_6^B \right)q_6 + 
 \left( \frac{\partial}{\partial B_y}  \lambda_6^C \right)r_6   
 \right],  \\
  \frac{d r_6}{ dt} =  -r_6\left [ \left( \frac{\partial}{\partial B_y}  \lambda_6^A \right)p_6 +
 \left( \frac{\partial}{\partial B_y}  \lambda_6^B \right)q_6 + 
 \left( \frac{\partial}{\partial B_y}  \lambda_6^C \right)r_6   
 \right],  \\  
 \frac{d s_6}{ dt} =  -s_6\left [ \left( \frac{\partial}{\partial B_y}  \lambda_6^A \right)p_6 +
 \left( \frac{\partial}{\partial B_y}  \lambda_6^B \right)q_6 + 
 \left( \frac{\partial}{\partial B_y}  \lambda_6^C \right)r_6   
 \right],  \\   $ \\

$ \frac{dx}{ dt} = \lambda_7^A, \>\> \frac{dy}{ dt} = \lambda_7^B,  
\>\> \frac{dz}{ dt} =\lambda_7^C  , \>\>  
\frac{d B_z}{ dt} = p_7\lambda_7^A + q_7\lambda_7^B + 
r_7\lambda_7^C + s_7,  \\ 
 \frac{d p_7}{ dt} = -p_7\left [ \left( \frac{\partial}{\partial B_z}  \lambda_7^A \right)p_7 +
 \left( \frac{\partial}{\partial B_z}  \lambda_7^B \right)q_7 + 
 \left( \frac{\partial}{\partial B_z}  \lambda_7^C \right)r_7   
 \right], \\
  \frac{d q_7}{ dt} = -q_7\left [ \left( \frac{\partial}{\partial B_z}  \lambda_7^A \right)p_7 +
 \left( \frac{\partial}{\partial B_z}  \lambda_7^B \right)q_7 + 
 \left( \frac{\partial}{\partial B_z}  \lambda_7^C \right)r_7   
 \right],  \\
  \frac{d r_7}{ dt} =  -r_7\left [ \left( \frac{\partial}{\partial B_z}  \lambda_7^A \right)p_7 +
 \left( \frac{\partial}{\partial B_z}  \lambda_7^B \right)q_7 + 
 \left( \frac{\partial}{\partial B_z}  \lambda_7^C \right)r_7   
 \right],  \\  
 \frac{d s_7}{ dt} =  -s_7\left [ \left( \frac{\partial}{\partial B_z}  \lambda_7^A \right)p_7 +
 \left( \frac{\partial}{\partial B_z}  \lambda_7^B \right)q_7 + 
 \left( \frac{\partial}{\partial B_z}  \lambda_7^C \right)r_7   
 \right], $ \\

$ \frac{dx}{ dt} = \lambda_8^A, \>\> \frac{dy}{ dt} = \lambda_8^B,  
\>\> \frac{dz}{ dt} =\lambda_8^C  , \>\>  
\frac{d \pi}{ dt} = p_8\lambda_8^A + q_8\lambda_8^B + 
r_8\lambda_8^C + s_8,  \\
 \frac{d p_8}{ dt} = -p_8\left [ \left( \frac{\partial}{\partial  \pi}  \lambda_8^A \right)p_8 +
 \left( \frac{\partial}{\partial \pi}  \lambda_8^B \right)q_8 + 
 \left( \frac{\partial}{\partial \pi}  \lambda_8^C \right)r_8   
 \right], \\
  \frac{d q_8}{ dt} = -q_8\left [ \left( \frac{\partial}{\partial \pi}  \lambda_8^A \right)p_8 +
 \left( \frac{\partial}{\partial \pi}  \lambda_8^B \right)q_8 + 
 \left( \frac{\partial}{\partial \pi}  \lambda_8^C \right)r_8   
 \right],  \\
  \frac{d r_8}{ dt} =  -r_8\left [ \left( \frac{\partial}{\partial \pi}  \lambda_8^A \right)p_8 +
 \left( \frac{\partial}{\partial \pi}  \lambda_8^B \right)q_8 + 
 \left( \frac{\partial}{\partial \pi}  \lambda_8^C \right)r_8   
 \right],  \\  
 \frac{d s_8}{ dt} =  -s_8\left [ \left( \frac{\partial}{\partial \pi}  \lambda_8^A \right)p_8 +
 \left( \frac{\partial}{\partial \pi}  \lambda_8^B \right)q_8 + 
 \left( \frac{\partial}{\partial \pi}  \lambda_8^C \right)r_8     
 \right], \\ $ \\ 

to avoid too lengthy formulas we used the standard eigenvalue notations \\
($\lambda^{j}_{k},  \>\> j=A,B,C \>\> k=5,6,7,8 $) for the last four magneto-acoustic waves.  
The complete system of equations for the characteristic curves are $8 \times 8 = 64$ 
equations, unfortunately the author has problems to interpret the results in the present form. 
We can restrict the motion to reduce the complexity of the dynamics. 
The simplest case is a one dimensional motion  in a one dimensional magnetic 
field which means a system with 4 PDEs. A possible stability analysis of a restricted motion 
may give new insight into the dynamics of the MHD equations. We can imagine that even 
more complex analysis like the Panlev\'e test or some kind of stability analysis can be applied to the 
presented differential equation system, unfortunately 
such investigation are out of the skill of the author. We can only hope that our work 
may give some impetus and motivate some reader to start such studies and investigate the 
equations presented above. 

\section{Summary}
We briefly presented how the generalized characteristics of the homogenous magneto hydrodynamical 
equations can be derived. Unfortunately, no deeper e.g. geometrical interpretation of the equation 
could be found, other detailed analysis is also lacking or may remain to the reader.

%%%%%%%%%%%%%%%%%%%%%%%%%%%%%%%%%%%%%%%%%%%%%%%%%%%%%%%%%%%%%%%%%%%%%%                                    

%%%%%%%%%%%%%%%%%%%%%%%%%%%%%%%%%%%%%%%%%%%%%%%%%%%%%%%%%%%%%%%%%%%%%%%                                    
%\end{multicols}
%%%%%%%%%%%%%%%%%%%%%%%%%%%%%%%%%%%%%%%%%%%%%%%%%%%%%%%%%%%%%%%%%%%%%%%                                    


\begin{references}
%\vspace*{-15mm}
%\bibitem[*]{present} 
%present address: 

\bibitem{powell} Kenneth G. Powell \\
{\it{An approximate Riemann solver for Magnetohydrodynamics 
Technical Report TR-94-24 1994}}, 
Department of Aerospace Engineering, The University of Michigan 


\bibitem{difk} Courant R. and Hilbert D.  {\it{Methods of Mathematical Physics 
}} Vol. II Wiley-Interscience New York, 1962                                       

\bibitem{wendr} Stewart H. B. and Wendroff B., J. Comp. Phys. 
{\bf{56}}, 363, (1984)                               


\bibitem{arik} Arik Melikyan \\
 {\it{ Generalized Characteristics of first order 
PDEs}} Birkh\"auser, 1998                                       

\bibitem{daferm}  Dafermos  C.M.\\
Generalized Characteristics and the Structure of Solutions of Hyperbolic Conservation Laws
 Indiana University Mathematics Journal {\bf{26}},                                      
1097, (1997)                                                                                               

\bibitem{daferm2} C.M. Dafermos, X. Geng \\ 
Generalized characteristics uniqueness and regularity of solutions in a hyperbolic system 
of conservation laws \\
Annales de l'I.H.P., section C, tome 8, $n^0$' 3-4 (1991), p. 231-269. \\           
http://www.nudam.org/item?id=AIHPC$_-$1991$_{--}$8$_{-}$3-4$_{-}$231$_{-}$0                                           

\end{references}
\end{document}